\documentclass{blois}
\pdfoutput=1 
\usepackage{jinstpub} 
\usepackage{HGTD_TDR-defs}
\usepackage{multirow}
\usepackage{rotating}
\usepackage{mathrsfs}
\usepackage{booktabs}
\usepackage{longtable}
\usepackage{placeins}
\usepackage{amssymb}
\usepackage{upgreek}
\usepackage{svg}
\usepackage{comment}
\usepackage{pdfpages}
\usepackage{siunitx}
\usepackage{lineno}
\usepackage{float}
\usepackage{lineno}

\notoc

\title{\boldmath Inter-pad dead regions of irradiated FBK Low Gain Avalanche Detectors}

\author[1]{B. Darby,\footnote{Corresponding author.}}
\author[1]{S.M. Mazza}
\author[1]{F.~McKinney-Martinez}
\author[1]{R.~Padilla}
\author[1]{H.~F.-W.~Sadrozinski}
\author[1]{A.~Seiden}
\author[1]{B.~Schumm}
\author[1]{M.~Wilder}
\author[1]{Y.~Zhao}

\author[2,6]{R.~Arcidiacono}
\author[2]{N.~Cartiglia}
\author[2]{M.~Ferrero}
\author[2]{M.~Mandurrino}
\author[2,5]{V.~Sola}
\author[2]{A.~Staiano}

\author[3]{V.Cindro}
\author[3]{G.~Kranberger}
\author[3]{I.~Mandiz}
\author[3]{M.~Mikuz}
\author[3]{M.~Zavtranik}

\author[4,7]{M.~Boscardin}
\author[7,8]{G.F.~Della~Betta}
\author[4,7]{F.~Ficorella}
\author[7,8]{L.~Pancheri}
\author[4,7]{G.~Paternoster}

\affiliation[1]{SCIPP, University of California Santa Cruz, 1156 high street, Santa Cruz (CA), US}
\affiliation[2]{INFN Torino, Torino (TO), Italy}
\affiliation[5]{Dipartimento di fisica, Universita' di Torino, Via Pietro Giuria 1, Torino (TO), Italy}
\affiliation[6]{Universita' del Piemonte Orientale, Via Teresa Michel 11, Alessandria (AL), Italy}
\affiliation[3]{JSI and Department of Physics, University of Ljubljana, Jamova cesta 39, Ljubljana, Slovenia}
\affiliation[4]{Fondazione Bruno Kessler, Via Sommardive 18, Povo (TN), Italy}
\affiliation[7]{TIPFA-INFN, Via Sommardive 14, Povo (TN), Italy}
\affiliation[8]{Dipartimento di Fisica, Universita' di Trento, Via Sommardive 14, Povo (TN), Italy}

\emailAdd{bdarby@ucsc.edu}

\abstract{Low Gain Avalanche Detectors (LGADs) are a type of thin silicon detector with a highly doped gain layer. LGADs manufactured by Fondazione Bruno Kessler (FBK) were tested before and after irradiation with neutrons. In this study, the Inter-pad dead regions (IPDRs), defined as the width of the distances between pads, were measured with a TCT laser system. The response of the laser was tuned using $\beta$-particles from a $^{90}$Sr source. These insensitive "dead zones" are created by a protection structure to avoid breakdown, the Junction Termination Extension (JTE), which separates the pads. The effect of neutron radiation damage at \fluence{1.5}{15}, and \fluence{2.5}{15} on IPDRs was studied. These distances are compared to the nominal distances given from the vendor, it was found that at the high fluence there
is a better matching of the insensitive region to the designed inter-pad region.}


\arxivnumber{2111.12656} 

\proceeding{The 12th international conference on position sensitive detectors\\
Sunday 12 - Friday 17 September 2021 \\
University of Birmingham}

\begin{document}

\maketitle
\flushbottom

\section{Introduction}
\label{sec:intro}

Low Gain Avalanche Detectors (LGADs) are thin silicon sensors with modest internal gain and exceptional time resolution~\cite{bib:LGAD,bib:MarTorino,bib:UFSD300umTB}. The internal gain is due to a highly doped p+ region (called multiplication or gain layer) just below the n-type implants of the electrodes.
Sensors in this study were produced by Fondazione Bruno Kessler (FBK) and are part of the UFSD-3.2 production for the ATLAS~\cite{CERN-LHCC-2020-007} and CMS~\cite{CMS:2667167} timing layers for HL-LHC (details on the production can be found in \cite{TIPPradhard}), in particular in this study sensors from Wafer 19 are studied which have a deep carbonated gain layer with optimized carbon dose and a bulk thickness of 45um. 
The arrays used in this study have a nominal inter-pad distance (IP) value of 49~um which is the most stable inter-pad design from the UFSD-3.2 production. 
Traditional LGADs arrays need a protection structure, called Junction Termination Extension (JTE), between pads to avoid breakdown caused by the high electric field at the edge of the multiplication layer. 
Charges collected by the JTE do not pass through the multiplication layer, therefore in the region in between pads no gain is present.
The size of the insensitive region between pads, the so-called inter-pad dead region (IPDR), is here measured using a 1064~nm focused laser. 
Sensor arrays from FBK UFSD-3.2 have an opening in between pads that allows laser studies of the inter-pad region.
Furthermore the collected charge and time resolution are studied in the inter-pad region.
The sensors are studied before and after neutron irradiation to study the effect of radiation damage to the IPDR. Sensors were irradiated with 1~MeV neutrons at the TRIGA reactor in Ljubljana, the uncertainty on the fluence is around 5\%.

\section{Experimental setup}

The experimental set up consists of a laser Trans-Current-Technique (TCT) system and a beta source setup (called beta scope).
The laser TCT setup is as follows: sensors are mounted on fast analog amplifier boards (16 channels) with 1~GHz of bandwidth (designed at FNAL) and the board is read out by a fast oscilloscope (2GHz, 20Gs). The laser is focused on the sensor plane with a beam spot of roughly 20~$\mu m$ and has a wavelength 1064~nm (IR). The IR laser penetrates through the sensor and causes a linearly distributed ionization, the power of the laser is adjusted to mimic the response of a Minimum Ionizing Particle (MiP) . 
The board is on X-Y moving stages so the response of the sensor as a function of position can be evaluated.

For the beta scope, sensors are mounted in an enclosure which is cooled to -30C. The sensor is connected to a fast analog electronic board (up to 2 GHz bandwidth) digitized by a GHz bandwidth oscilloscope. A second board is also mounded with a known fast LGAD to act as a trigger and as time reference. Both of these boards are then exposed to beta particles by a $^{90}$Sr beta-source. Further explanation of the beta scope setup and analysis can be found in \cite{Mazza:2018jiz,bib:HPKirradiation35vs50,bib:HPKirradiationGalloway}.

\section{Data analysis}

\begin{figure}[htbp]
\centering 
\qquad
\includegraphics[width=0.47\textwidth]{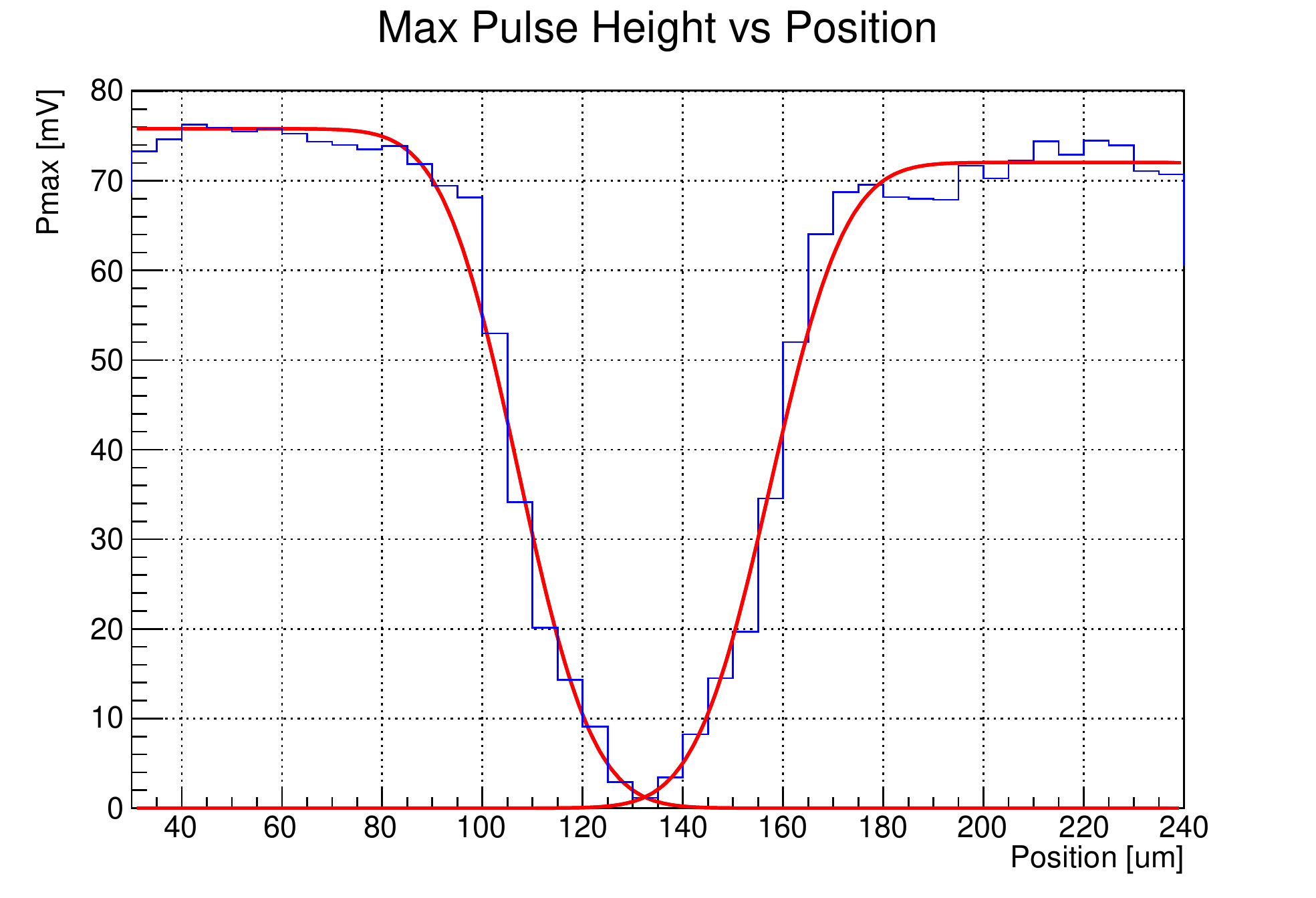}
\includegraphics[width=0.47\textwidth]{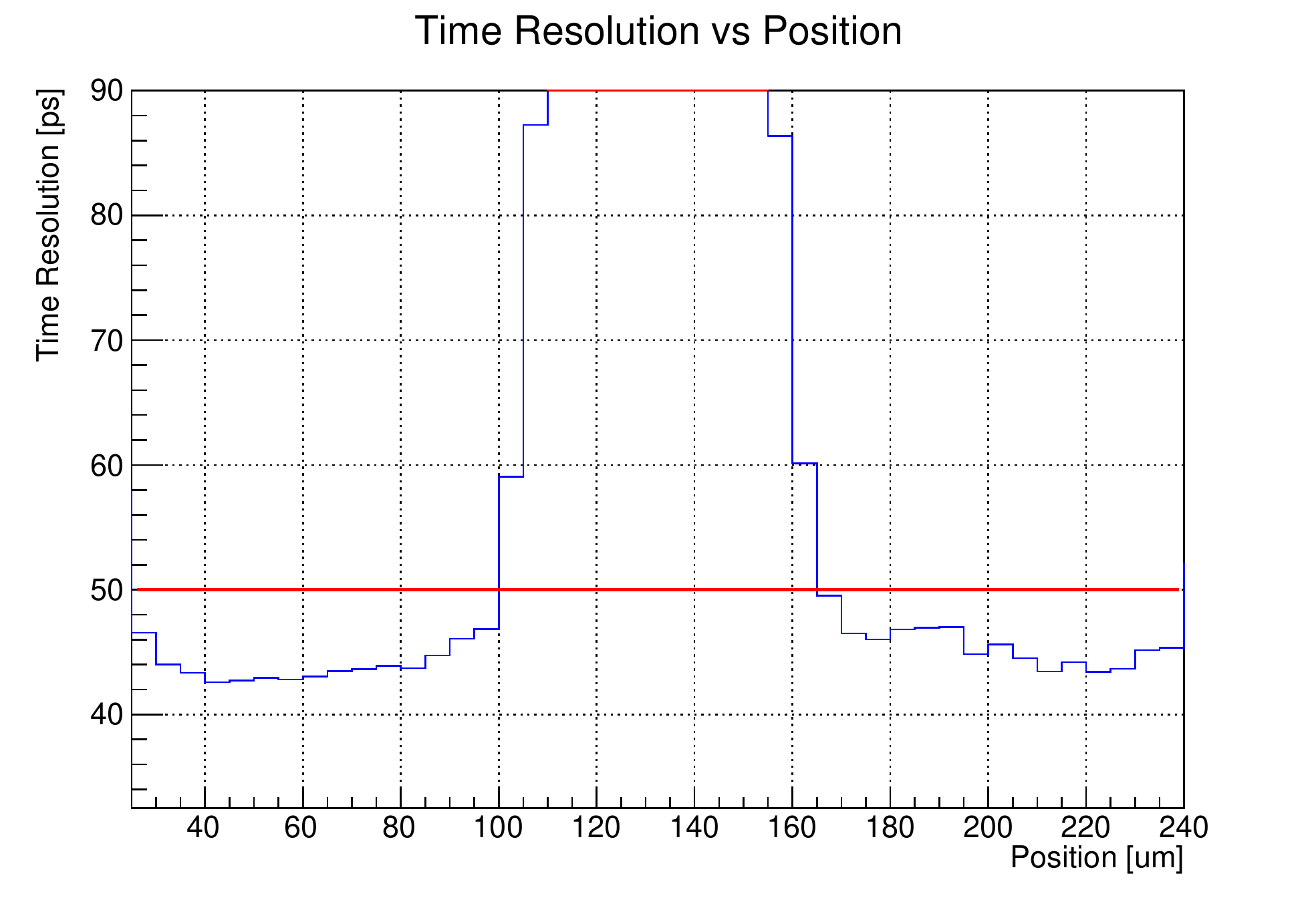}
\caption{Left: Pulse maximum vs Position of a FBK UFSD-3.2 detector irradiated to \fluence{2.5}{15} with a nominal IPDR of 49um. Measurements taken with a TCT laser. Right: Time resolution vs Position of a FBK UFSD-3.2 detector irradiated to \fluence{2.5}{15}.}
\label{fig:profile}
\end{figure}

The IPDR is measured in the TCT setup as the spatial distance in between the points were the pulse maximum ($P_{max}$) is half of the response in the center of the pads, this value is also called the 50-50 distance. An example of the scan in between pads for an FBK UFSD-3.2 sensor irradiated to \fluence{1.5}{15} is show in Figure~\ref{fig:profile}, the distance in the plot is roughly 50~um (10 steps of 5~um).
Data for the IPDR measurement is taken at several applied bias voltages.

The laser data was calibrated by adjusting the laser power to have the same Pmax response in the center of the pad as the one from the beta scope setup.
The collected charge and time resolution distributions as a function of Pmax is then extracted from the beta scope data and used to convert the laser Pmax profiles vs position to charge and time resolution profiles.

To evaluate the 50-50 distance, the $P_{max}$ distributions are fitted with a step function and the 50\% points of the fit is calculated.
Another way of defining IPDR is to measure the distance where a specific collected charge is satisfied. 
The collected charge is a critical parameter for readout electronics since it influences the efficiency of the response and the Jitter component of the time resolution, in case of the ATLAS HGTD readout chip (ALTIROC) the minimum collected charge to achieve 100\% response efficiency is 2.5~fC and to achieve a good Jitter is 4~fC.
The time resolution of the sensor was also calculated as a function of position combining the information from the beta scope data with the laser data.

\section{Results}
\begin{figure}[htbp]
\centering
\includegraphics[width=0.49\textwidth]{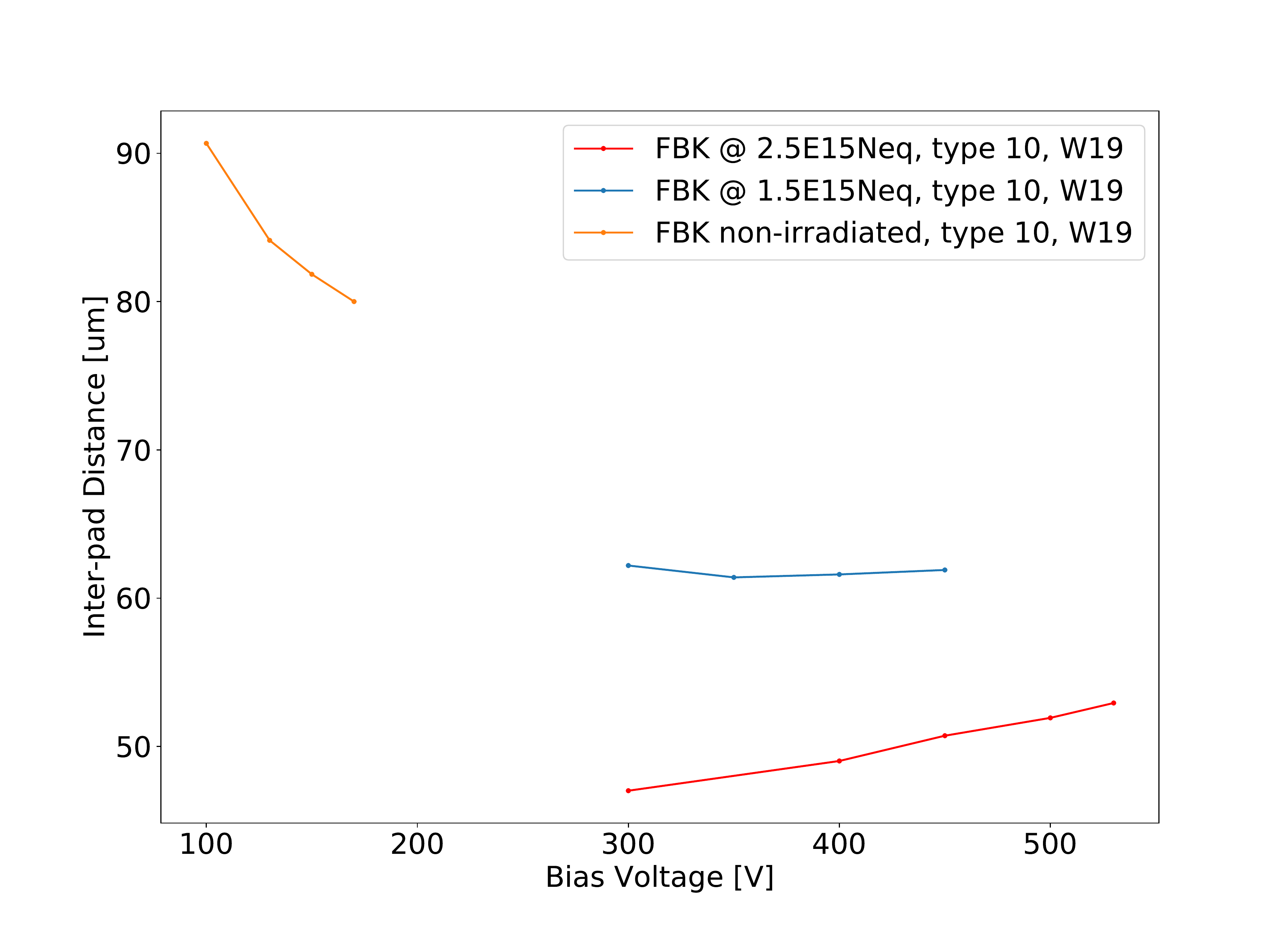}
\includegraphics[width=0.49\textwidth]{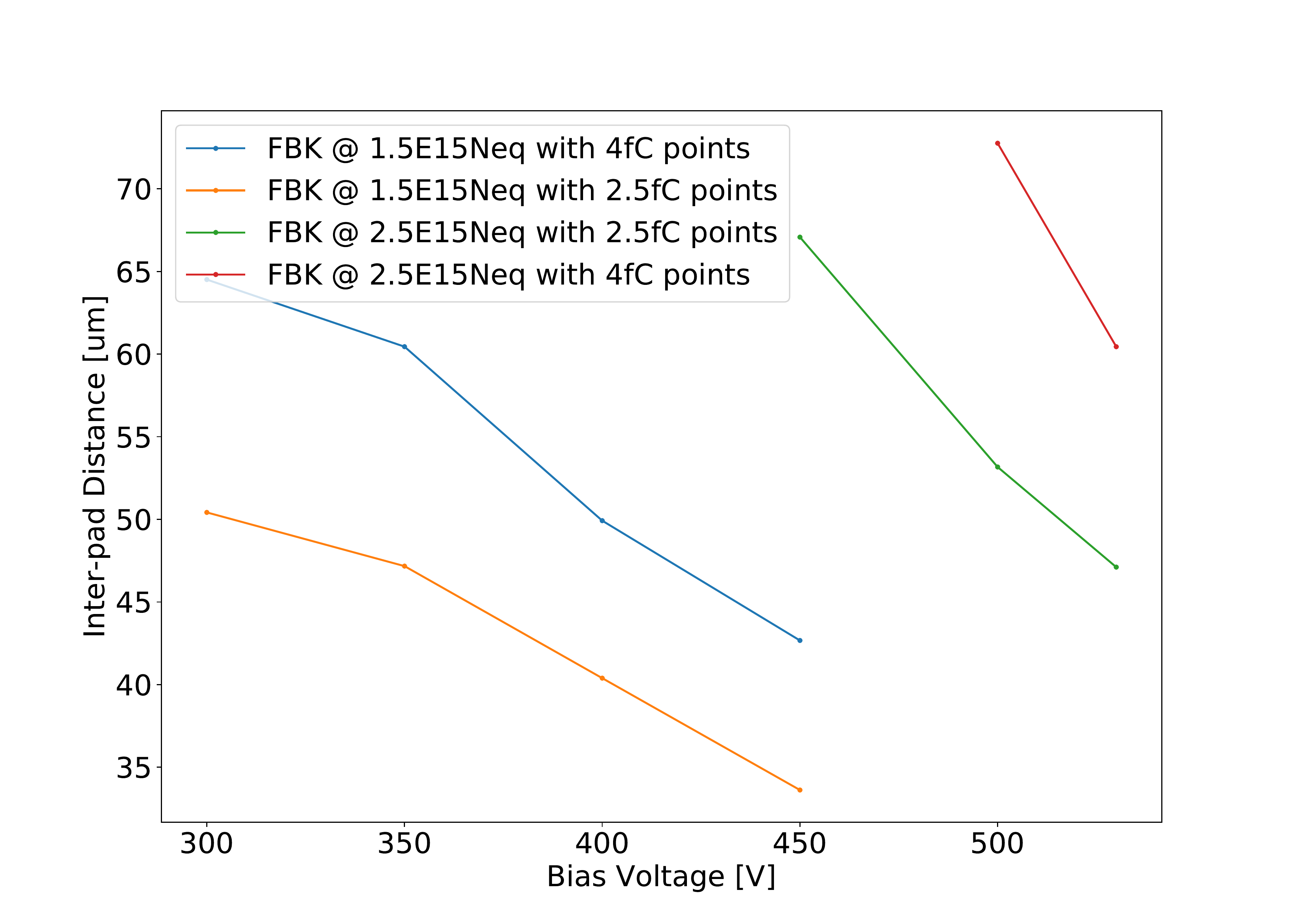}
\caption{Left: IPDRs of irradiated FBK detectors. The IPDR decreases to better match the nominal IPDR value of 49um. 
Right: IPDR vs Bias Voltage of irradiated detectors. Measuring the inter-pad distance using the charge collection on the detector at 2.5~fC and 4~fC. This method of using charge collection for measuring the IPDR does not work for the non-irradiated sensor due to the high gain.}
\label{fig:IPDRvolt}
\end{figure}

The 50-50 IPDR as a function of bias voltage for FBK UFSD-3.2 arrays is shown in Figure~\ref{fig:IPDRvolt}, Left. The nominal IPDR for this type of array is 50~um, it can be seen that before irradiation the actual 50-50 IPDR is around 80~um. However after irradiation the measured IPDR is closer to the nominal value. 
The distance between positions at which the pad surpasses 2.5~fC and 4~fC of collected charge is shown in Figure~\ref{fig:IPDRvolt}, Right. The values are shown only for irradiated sensors since before irradiation the gain is too high and the collected charge cannot be evaluated at low values. The IPDR for 2.5~fC (necessary for 100~\% response efficiency of the ATLAS readout chip) is 50~um or less, the IPDR for 4~fC (necessary for good Jitter response of the ATLAS readout chip) is 60~um or less.
In Table~\ref{leg:IPDRs} the measured IPDRs for all sensors is reported for all methods including the IPDR to achieve 50 ps of time resolution.

\begin{table}
\caption{IPDR and bias voltage by method: 50-50 and to reach collected charge of 2.5/4 fC and time resolution ($\sigma_T$) of 50 ps. FBK UFSD3.2 W19 type 10 arrays irradiated at fluence N[\fluence{}{15}].}
\label{leg:IPDRs}
\smallskip

\parbox{.49\textwidth}{
\begin{tabular}{|l|r|c|c|}

\hline
Method & Bias V. [V] & IPDR [um] & Fluence \\
\hline



50-50& 100 & 90.67 & 0 \\
& 130 & 84.13 & 0 \\
& 150 & 81.84 & 0 \\
& 170 & 80.00 & 0 \\
\hline
$\sigma_T$& 100 & 100.48 & 0 \\
& 130 & 87.86 & 0 \\
& 150 & 83.10 & 0 \\
& 170 & 83.49 & 0 \\

\hline
\hline

50-50& 300 & 46.9 & 2.5 \\  
& 400 & 48.9 & 2.5 \\
& 450 & 50.6 & 2.5 \\
& 500 & 51.8 & 2.5  \\
& 530 & 52.8 & 2.5  \\
\hline

$\sigma_T$& 450 & 70.48 & 2.5 \\
& 500 & 56.96 & 2.5 \\
& 530 & 52.05 & 2.5 \\

\hline
2.5fC& 450 & 66.92 & 2.5 \\
& 500 & 53.04 & 2.5 \\
& 530 & 46.98 & 2.5 \\
\hline

4fC& 500 & 72.59 & 2.5 \\
& 530 & 60.30 & 2.5 \\

\hline

\hline
\end{tabular}
}
\hfill
\parbox{.49\textwidth}{
\begin{tabular}{|l|r|c|c|}
\hline
Method & Bias V. [V] & IPDR [um] & Fluence \\
\hline

50-50 & 300 & 62.2 & 1.5\\
& 350 & 61.4 & 1.5  \\
& 400 & 61.6 & 1.5  \\
& 450 & 61.9 & 1.5 \\

\hline

$\sigma_T$& 300 & 73.54 & 1.5 \\
& 350 & 59.27 & 1.5 \\
& 400 & 53.13 & 1.5 \\
& 450 & 48.37 & 1.5 \\

\hline
2.5fC & 300 & 50.42 & 1.5\\
& 350 & 47.17 & 1.5\\
& 400 & 40.39 & 1.5\\
& 450 & 33.62 & 1.5\\


\hline

4fC & 300 & 64.51 & 1.5\\
& 350 & 60.45 & 1.5\\
& 400 & 49.92 & 1.5\\
& 450 & 42.67 & 1.5\\
\hline
\end{tabular}
}

\end{table}

\section{Conclusion}
The inter-pad distances (IPDR) of FBK detectors were measured before and after irradiation of neutrons at \fluence{1.5}{15} and \fluence{2.5}{15}. 
Before irradiation the measured 50-50 IPDR for FBK UFSD-3.2 W19 Type 10 array is higher than the nominal value given by the vendor (80~um instead of the nominal 49~um), a possible explanation is that the applied bias voltage is not enough to have straight field lines near the JTE. 
This would cause charges deposited under the gain layer to drift in the JTE region and not being multiplied, effectively increasing the no-gain region in between pads. 
However after radiation damage the 50-50 IPDR is more in agreement with the nominal value, since the applied bias voltage is increased after irradiation the field lines would be straighter. Therefore the results are in line with the given explanation.


{
\small
\acknowledgments
This work was supported by the United States Department of Energy, grant DE-FG02-04ER41286, and partially performed within the CERN RD50 collaboration.
Part of this work has been financed by the European Union's Horizon 2020 Research and Innovation funding program, under Grant Agreement no. 654168 (AIDA-2020) and Grant Agreement no. 669529 (ERC UFSD669529), and by the Italian Ministero degli Affari Esteri and INFN Gruppo V.
\bibliography{bib/TechnicalProposal,bib/hpk_fbk_paper,bib/HGTD_TDR}
}

\end{document}